\documentclass[pra,twocolumn,nofootinbib,eqsecnum,floatfix,superscriptaddress]{revtex4}
\usepackage{epsfig}
\usepackage{bm}
\usepackage{amsmath}
\usepackage{ulem}
\usepackage{datetime}
\input epsf
\numberwithin{equation}{section}
\renewcommand{\theequation}{\arabic{section}.\arabic{equation}}

\def\hP{\hat P}

\def\range{$[0,1]$}
\def\Re{{\rm Re}}

\def\EP{\text{EPE~}}
\def\t0{t_0}
\def\tf{t_f}
\def\Psih{{\hat\Psi}}
\def\Ch{{\hat C}}

\def\Sc{{\cal S}}

\begin{document}

\title{Decoherent Histories Quantum Mechanics with \\ One ``Real'' Fine-Grained History}

\author{Murray Gell-Mann}

\email{mgm@santafe.edu}

\affiliation{Santa Fe Institute, Santa Fe, NM 87501}

\author{James B.~Hartle}

\email{hartle@physics.ucsb.edu}

\affiliation{Santa Fe Institute, Santa Fe, NM 87501}
\affiliation{Department of Physics, University of California,
 Santa Barbara, CA 93106-9530}

\date{\today } 

\begin{abstract}
Decoherent histories quantum theory is reformulated with the assumption that there is one ``real'' fine-grained history, specified in a preferred complete set of sum-over-histories variables.  This real history is described by embedding it in an ensemble of comparable imagined fine-grained histories, not unlike the familiar ensemble of statistical mechanics.  These histories are assigned extended probabilities, which can sometimes be negative or greater than one. As we will show, this construction implies that the  real history  is not completely  accessible to experimental or other observational discovery. However, sufficiently and appropriately coarse-grained sets of alternative histories have standard probabilities providing information about the real fine-grained history that can be compared with observation. We recover the probabilities of decoherent histories quantum mechanics for sets of histories that are recorded and therefore decohere.  Quantum mechanics can  be viewed as a classical stochastic theory of histories with extended probabilities and a well-defined notion of reality common to  all decoherent sets of alternative  coarse-grained histories.
\end{abstract}



\maketitle

\section{Introduction}
\label{intro}

Decoherent histories quantum mechanics\footnote{For broad accounts of decoherent histories and related formulations, with references to important earlier literature, see  e.g. \cite{Gri02,Omn94,Gel94}. For a short tutorial see \cite{Har93a}.} (DH) is logically consistent, in agreement with experiment as far as is known, applicable to cosmology, consistent with the rest of modern physics including special relativity and quantum field theory, and generalizable to include quantum gravity\footnote{For the extension to quantum spacetime see e.g. \cite{Har95c}.}. It is a framework for quantum cosmology and for understanding large scale features of the quantum universe ranging from the approximate applicability of classical physics under suitable conditions to the number of e-foldings of inflation. It includes the Copenhagen quantum theory of laboratory experiment as an approximation adequate for measurement situations. It is the only presently available formulation of quantum theory with  all these properties. 

DH assigns probabilities to the members of decoherent sets of alternative coarse-grained histories of the universe.  By decoherent we mean that, as a consequence of the quantum state and dynamics of the universe,  there is negligible quantum interference between coarse-grained histories in the set.   A coarse-grained history can be regarded as a class of fine-grained ones. Fine-grained histories describe the system as completely as possible.  Feynman paths for a particle are an example. Decoherent sets of alternative coarse-grained histories are called realms.  

As usually formulated, DH presents two obstacles to the idea that there is  one unique, real, fine-grained history of the universe that we experience at a highly coarse-grained level. The first is that,  except for trivial cases, there are no decoherent sets of completely fine-grained histories.  Negligible interference requires coarse-graining such that the phases between histories are washed out.  The second obstacle is that there are realms that are mutually incompatible in the sense that there is no common finer-grained realm of which they are both coarse-grainings. There appears to be no connection between the real history of one realm and that of an  incompatible one.

In this paper we show how to overcome these obstacles. 
We overcome the first one by extending the notion of probability to include values outside the range \range. Extended probabilities can then be assigned to any set of alternative histories, in particular to fine-grained ones. We overcome the second obstacle by restricting the allowed sets of histories to those describable in a  preferred set of variables --- those used in a sum-over-histories treatment of quantum theory. The result will be a formulation of quantum theory that at first sight appears to be  both an extension and a contraction of DH. It is an extension in that it uses extended probabilities; it is a contraction in the sense of using only a preferred set of variables. However, we shall see that in realistic situations it is equivalent to a sum-over-histories formulation of DH.  

There are two basic starting points for  this formulation. The first is the ensemble method of J. Willard Gibbs \cite{Gib1902}, which has proved to be essential for describing the coarse-grained regularities of physical systems about which we have little fine-grained information, as in statistical mechanics. The same method is also useful for characterizing the complexity of such regularities  \cite{GL04}. The second starting point is the notion of extended probability \cite{Har08}. 

The statistical mechanics of a classical gas of $N$ particles in a box illustrates the ensemble method.  The gas is described at a moment of time $t_0$ by giving the positions and momenta of the $N$ particles, that is, by giving a point, $z_0=z(t_0)$, in the $6N$ dimensional phase space of the system. This point in phase space evolves in time; $z(t)$ follows Newton's deterministic laws. That is the real fine-grained history of the gas.  The evolution by a deterministic law  is a regularity of the fine-grained history, but not one that is completely accessible or useful to us when $N$ is large. There is no practical hope of measuring, storing, retrieving, or computing with all the information involved in describing these regularities. The accessible, useful regularities, such as those summarized by the Navier-Stokes equation, are very coarse-grained.  To describe them we conceptually embed the one real history of the particles in an ensemble (a set with probabilities) of imagined comparable fine-grained histories with various different initial conditions. The term ensemble indicates that probabilities are assigned to these initial conditions  in such a way that the coarse-grained regularities have high probability. For example, one might consider a time-dependent coarse graining based on hydrodynamic quantities with probabilities representing local equilibrium. The phenomenological equations of classical hydrodynamics then hold with high probability\footnote{See, e.g. \cite{GH07} for an  exposition in the present context.}.
 The assumption that the one real history is typical in this ensemble allows us to use the probabilities to bet on what will really happen in the future and what really happened in the past. 

We will show how the quantum mechanics of a closed system can be formulated in a similar way. We assume the preferred variables of sum-over-histories quantum mechanics are the ones to be used  for describing fine-grained histories\footnote{This assumption fits with the idea that the sum-over histories formulation of quantum theory may be a more general and therefore a more fundamental framework for quantum mechanics \cite{Gel88,Har92}. Extensions of usual quantum theory to incorporate quantum spacetime seem naturally formulated in this way (e.g \cite{Har07,Har95c}).}. The one real fine-grained history of the universe is embedded conceptually in an ensemble of alternative fine-grained histories, each of which is assigned an {\it extended} probability \cite{Har08} based on an assumed quantum state and Hamiltonian for the system.  Extended probabilities obey the usual rules of probability theory except that they can  be negative or greater than one for an alternative for which it cannot be determined whether it occurs or not  (as on the alternative histories in the two-slit experiment).   From the extended probabilities of  fine-grained histories, extended probabilities can be constructed for sets of coarse-grained alternative histories that are classes of fine-grained histories.  Sufficiently and appropriately  coarse-grained ensembles have only probabilities between $0$ and $1$  and can therefore  be used to bet on the outcomes of experiments that test the theory.  The result, as mentioned in \cite{Har08},  is that quantum mechanics can be viewed as a classical stochastic theory with extended probabilities.

This formulation of the quantum mechanics of closed systems is not a replacement for 
decoherent histories quantum theory but rather a different starting point for it. As we will see, we recover a sum-over-histories version of  DH in the end for realistic situations. 

Our paper is organized as follows:   Section \ref{histens} explains how quantum mechanics can be formulated as a prescription for the extended probabilities of an ensemble of fine-grained histories containing one real history.  Section \ref{betsrecords} discusses the role of records and the connection to decoherent histories quantum theory. Section \ref{implications} discusses the implications of this formulation of quantum mechanics.  Section \ref{conclusion} contains a summary and further discussion of the implications.

\section{The Quantum Ensemble of Fine-Grained Histories}
\label{histens}
This  section formulates the quantum mechanics of a closed system as a theory of one real fine-grained history embedded in an ensemble of comparable alternative fine-grained histories described in a preferred set of variables and  assigned extended probabilities. To keep the discussion manageable, we neglect gross fluctuations in the geometry of spacetime. We then in effect consider a system of a box of particles or fields moving in a fixed, presumably  expanding, background spacetime. Well-defined notions of space and time are then available, as well as the usual machinery of quantum theory ---  amplitudes, operators, a Hilbert space of states, etc. The important thing is that the system is closed so that observers and measuring apparatus (if any) and  all other physical systems are within the box and are part of the quantum system being described. We can think of this as a simplified model of the universe. The fundamental theory of this closed system consists of two parts: the system's Hamiltonian $H$ specifying the dynamics (assumed independent of time for simplicity) and the quantum state specified by a wave function $\Psih$.

\subsection{Four ingredients of the formulation}
This formulation is specified by four ingredients:
\begin{itemize}

\item{} The preferred set of variables in terms of  which the one real fine-grained history is described, as well as the alternative fine-grained histories of the ensemble in which the one history is embedded. 

\item{} An extended notion of probability that reflects the notion of ignorance in quantum mechanics. 

\item{} The prescription that assigns extended probabilities to the members of the ensemble of alternative fine-grained histories using  the system's quantum state and Hamiltonian.  

\item{}   Coarse graining of the fine-grained set of histories leading to coarse-grained sets that can be recorded and decoherent and can have standard probabilities. 

\end{itemize}
We now discuss these in turn:

{\it Preferred Variables:}  The fine-grained histories are described by a preferred set of variables which we take to be those of a sum-over-histories formulation of quantum mechanics. They are histories of particle positions in the case of particles, four-dimensional field configurations --- both bosonic and fermionic --- in the case of quantum field theory, and histories of geometries and fields in the case of semiclassical quantum gravity.   Histories of these variables are assumed to be the most refined description of the system possible. 

A strong case can be made that these histories are adequate for the prediction of all observable quantities \cite{FH65}. Particle momentum, for instance, can be defined in terms of histories of position in a time-of-flight setup that would measure velocity.  We do not rule out introducing operators, transformation theory, etc at some later stage but we begin by assuming this preferred set of variables for fine-grained histories.

To describe these preferred variables we assume a particular Lorentz frame and let $t$ be the time coordinate of that frame. 
We denote the preferred variables by $q^i$ or just $q$ for short. For particles $i$ might be $x,y,z$ and a particle label. For fields $i$ would include the label $\vec x$ of the spatial point. We denote the configuration space spanned by $q^i$ by $\cal C$. A fine-grained history is a path $q(t)$ in $\cal C$ that we assume to be single-valued --- one and only one value of $q$ for each $t$. The set of all fine-grained histories between an arbitrary pair of times  $t_0$ and $t_f$ is the set of all such paths $\{q^i(t)\}$ between these two times. They are continuous but typically non-differentiable. 

{\it Extended Probabilities:}
Probabilities can be usefully understood as instructions for making fair bets \cite{deF37}. To hold that the probability of an alternative $\alpha$ is $p(\alpha)$ means the following:  Suppose that there is a bet on whether $\alpha$ occurs with  payoff  $S_\alpha$ (of either sign) if it does. You will put up  $p(\alpha)S_\alpha$ and consider it a fair bet.  Probabilities can thus be said to express our ignorance with respect to whether the alternative $\alpha$ occurs. 

All the usual rules of probability theory, including the restriction of values to the range \range, follow from the requirement that a bookie not be able to offer you  a ``Dutch book'' in which you will put up the stake $p(\alpha)S_\alpha$ but be guaranteed to always lose, not just on average, but each time the bet is made \cite{deF37,Cavup}.

Implicit in the above definition of probability in terms of fair bets is the assumption that it can be settled whether the alternative $\alpha$  occurs or not.  Elementary physics assumes that any alternative that can be described can be determined  without significantly affecting its value. Every alternative in elementary physics is therefore in principle the basis of a settleable bet, however difficult it may be to settle it in practice. For instance, we assume that the value of the initial condition $z_0$ of the box of gas discussed in the preceding section is in principle the basis for a settleable bet even though it is impossible in practice  to determine $z_0$. Thus, when we construct an ensemble reflecting our ignorance of $z_0$,  we assign probabilities obeying the usual rules to the different values it might take. 
\begin{figure}
\begin{center}
\includegraphics[width=3in]{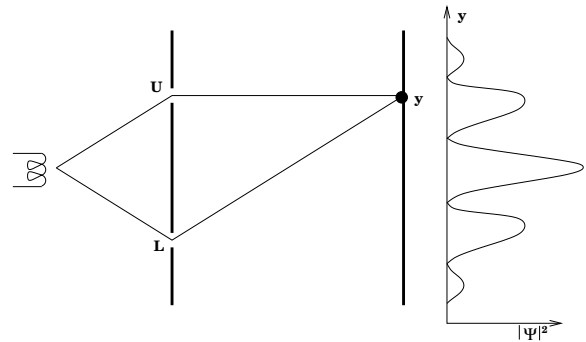}
\caption{The two-slit experiment. An electron gun at left emits an electron
traveling towards a screen with two slits, its progress in space recapitulating
its evolution in time. The electron is detected at a further screen at a position $y$ with a probability density that exhibits an interference pattern. A coarse-grained set of histories for the electron is defined by specifying the slit ($U$ or $L$) through which the electron passes through and ranges $\Delta$ of the position $y$ where it is detected. In the absence of the record of  a measurement it is not possible to settle a bet on the which of these histories occurred.}   
\label{2slit}
\end{center}
\end{figure}

But in quantum theory there are alternatives that can be described but are not the basis for settleable bets.  A classic example is provided by the two-slit experiment illustrated in Figure 1. A bet on whether the electron went through the upper slit or the lower slit is not settleable without carrying out a measurement that would significantly disturb the system. Alternatives that are not the basis of settleable bets are a new kind of ignorance not found in elementary physics and they can be usefully described by extending the classical notion of probability. In \cite{Har08} we proposed the following simple extension: Keep the rules of probability as they are, but allow the values to be outside the range \range\ for alternatives corresponding to non-settleable bets. Evidently this extension does not lead to Dutch books because no book of any kind can be made on non-settleable alternatives. 

Even one out-of-range probability in an exhaustive set of exclusive alternatives means that there are no settleable bets on which alternative occurs. If there were, a Dutch book could be constructed using only the alternatives that have out-of-range extended probabilities.
The excursion out of the normal range \range\ is a measure of how much coarse graining is needed to find alternatives that have probabilities in the normal range \cite{Har08,Har04}. 

The quantum history ensemble will consist of alternative fine-grained histories assigned extended probabilities. We now turn to how the ensemble is constructed. 

{\it The Fundamental Distribution:} 
To complete the formulation of quantum theory starting with an  ensemble of fine-grained histories, we need to specify the fundamental distribution $w[q(t)]$ that assigns an extended probability\footnote{The distribution is a probability functional density for continuous $q$, but we rely on the reader to make this qualification where appropriate.}  to each history $q(t)$. This formula will depend on the initial quantum state of the box represented by a Schr\"odinger picture wave function $\Psih(q,t_0)$. (Hats denote the Schr\"odinger picture.) This is just as much a part of the necessary theoretical structure for prediction in the universe as the Hamiltonian\footnote{In more general quantum gravitational contexts a theory of the quantum state like the one that yields  the no-boundary wave function \cite{HH83} may derive the state from the dynamical theory.}  $H$. There are no quantum-mechanical predictions of any kind that do not depend on both. The Schr\"odinger picture wave function evolves in time according to 
\begin{equation}
\label{Seqn}
\Psih(q,t) = e^{-iH(t-t_0)/\hbar} \Psih(q_0,t_0) .
\end{equation}

To keep the discussion manageable assume that all coarse-grained alternatives lie between an initial  $t_0$ and a final time $t_f$. That is not a loss of generality since $t_f$ can be as large as desired.   Elementary causality shows that we need not consider fine-grained histories at times later than $t_f$ (e.g \cite{GH93a}, Section VIIC). Thus, we consider single valued histories $q(t)$ that take initial values $q_0$ at $t=\t0$ and final values $q_f$ at $t=\tf$. We then postulate the fundamental distribution of extended probabilities is \cite{Har08}
\begin{equation}
\label{fund_dist}
w[q(t)] \equiv {\rm Re}\left[ \Psih^*(q_f,t_f) \exp\{i \Sc[q(t)]/\hbar\} \Psih(q_0, t_0) \right] \ ,
\end{equation}
where $\Sc[q(t)]$ is the action functional corresponding to the Hamiltonian $H$. We will show in Section \ref{betsrecords} that this formula reproduces the probabilities of DH  when there are records of histories that can be used to settle bets on which history actually occurs. At this point in the exposition, however, \eqref{fund_dist} is a postulate. 

So defined, the extended probabilities sum to one since the integral over all paths from $\t0$ to $\tf$ (including over $q_0$ and $q_f$) is 
\begin{equation}
\label{normalization}
\int \delta q \ w[q(t)] = \int dq_f \Psih^*(q_f,\tf) \Psih(q_f,\tf) = 1 
\end{equation}
because the sum over all paths just evolves the wave function at $t_0$ to $t_f$ to give \eqref{normalization}. 

 The distribution $w[q(t)]$ defined by \eqref{fund_dist} will have values outside the range \range~ for some $q(t)$. To see that, suppose there is a history $q(t)$ with $w[q(t]$  within the range \range. Then a variation of $q(t)$ that leaves $q_0$ and $q_f$ unchanged can still contribute significantly to the action in \eqref{fund_dist} and change the sign of $w[q(t)]$ from positive to negative. As we will see in Section \ref{betsrecords},  these negative extended probabilities mean that the set of fine-grained histories is not the basis for a settleable bet on what the real fine-grained history is like. 

{\it Coarse Graining.}
 The set of alternative fine-grained histories $\{q(t)\}$ can be coarse grained by partitioning it into an exhaustive set of exclusive classes $c_\alpha$ with $\alpha=1,2, \cdots$. The extended probability of the class $c_\alpha$ is the sum of the extended probabilities of all its members, that is
\begin{equation}
\label{cgprobs}
p(\alpha) = \int_{c_\alpha} \delta q  \ w[q(t)] \  ,
\end{equation}
where the sum is over all $q(t)$ in the class $c_\alpha$, including a sum over $q_0$ and $q_f$. 

 Sufficient coarse graining will lead to sets with all positive probabilities  if only because the completely coarse-grained set consisting of all histories has probability one\footnote{See \cite{Har08} for a more quantitative discussion of how coarse graining leads to usual probabilities.}. Suitably  coarse grained sets can be recorded and decoherent, and can have standard probabilities. They therefore describe alternatives that are the basis of settleable bets. 
We  discuss this in more detail in the next section, but we conclude this one with introducing useful operator representations of most of our formulae.

\subsection{Operators}

Using \eqref{fund_dist},  the expression \eqref{cgprobs} for the extended probability of a coarse-grained history $c_\alpha$ can be written more explicitly as
\begin{align}
p(\alpha)&=\int dq_f \int dq_0 \int_{[q_0 c_\alpha q_f] }\delta q \nonumber \\ &\times{\rm Re}\left[ \Psih^*(q_f,t_f) \exp\{i \Sc[q(t)]/\hbar\} \Psih(q_0, t_0) \right] \ ,
\label{cgpexplicit}
\end{align}
where the notation $[q_0 c_\alpha q_f]$ means that the sum is over all histories in the class $c_\alpha$ that begin at $q_0$ and end at $q_f$.
This can be conveniently written in operator form as 
\begin{equation}
\label{pfrmC}
p(\alpha)= \Re[\langle \Psih(t_f)|\Ch_\alpha | \Psih(t_0)\rangle] \ ,
\end{equation}
where the matrix elements of the class operator $\Ch_\alpha$ in the Schr\"odinger picture are defined by
\begin{equation} 
\label{pathforC}
 \langle q_f|\Ch_\alpha |q_0 \rangle   \equiv \int_{[q_0,c_\alpha,q_f]} \delta q e^{iS[q(t)]/\hbar}. 
\end{equation}
and  $\Psih(q,t)=\langle q |\Psih(t) \rangle$. 

The formulae simplify even further in the Heisenberg picture. Referred to the time $t_0$, the Heisenberg picture state is 
\begin{equation}
\label{Hstate}
|\Psi\rangle \equiv e^{iH(t_f-t_0)/\hbar }|\Psih(t_0)\rangle . 
\end{equation}
The equivalence between sum-over-histories evolution and Hamiltonian evolution shows that the Heisenberg picture state $|\Psi\rangle$ is constant in time. 
Then, if we define
\begin{equation}
\label{StoH}
C_\alpha \equiv e^{iH(t_f-t_0)/\hbar} \Ch_\alpha, 
\end{equation}
we find \footnote{This is the expression  used by Goldstein and Page \cite{GP95} to define their linear positivity condition that restricts sets of coarse-grained histories to those for which \eqref{funddistH} is positive. The fundamental distribution  \eqref{fund_dist} is a related expression that assigns  extended probabilities to  fine-grained histories without restriction.}
\begin{equation}
\label{funddistH}
p(\alpha)=\Re [ \langle\Psi|C_\alpha|\Psi\rangle]. 
\end{equation}

Extended probabilities satisfy all the rules of probability theory except the requirement that probabilities lie in the range \range. In particular, they must satisfy the usual sum rules connecting the extended probabilities for coarse-grained sets related by operations of fine and coarse graining. Suppose that $\{{\bar c}_{\bar \alpha}\}$ is a coarse graining of a set $\{c_\alpha\}$. This means that the even coarser-grained set is a partition of the finer-grained set into an exhaustive set of exclusive classes so that 
\begin{equation}
\label{cbarc}
{\bar c}_{\bar \alpha}=\cup_{\alpha \in {\bar\alpha}}\  c_\alpha \ . 
\end{equation}
The rules of extended probabilities require the sum rules
\begin{equation}
\label{psum}
{\bar p}({\bar\alpha})=\sum_{\alpha \in {\bar\alpha}} p(\alpha). 
\end{equation}
These are satisfied exactly as a consequence of \eqref{cgprobs}, which leads to 
\begin{equation}
\label{Csum}
{\bar C}_{\bar \alpha} =\sum_{\alpha \in {\bar\alpha}} C_\alpha .
\end{equation}
 
A familiar example of a coarse graining of a fine-grained set of histories concerns the motion of a particle in one dimension $x$. The set of fine-grained paths can be partitioned into classes by using  exhaustive sets of exclusive intervals of $x$, $\{\Delta^k_{\alpha_k}\}$, with $\alpha_k=1,2,\cdots$, at a sequence of times $t_k$ $k=1,2, \cdots, n$.  The index $k$ allows for the sets to be different at different times. The index $\alpha_k$ labels the interval in each set. The alternatives at each time correspond to an exhaustive set of exclusive (Schr\"odinger picture) projection operators $\{\hP^k_{\alpha_k}\}$ onto the intervals $\Delta^k_{\alpha_k}$. 
A coarse- grained history $\alpha$  is defined by a particular sequence of intervals at the possible times, $\alpha=(\alpha_1,\alpha_2,\cdots, \alpha_n)$. The class $c_\alpha$ consists of the bundle of paths $q(t)$ that pass through these intervals at the assigned times. The class operator in the Heisenberg picture defined by \eqref{pathforC} can be shown (e.g \cite{Cav86}) to be given by 
 \begin{equation}
 C_\alpha = P^n_{\alpha_n}(t_n)\cdots P^1_{\alpha_1}(t_1).
 \label{chain}
 \end{equation}
 The extended probabilities  of these histories are given by \eqref{funddistH}; some explicit calculations can be found in \cite{Har04}.

\section{Settleable Bets, Records, and Decoherence}
\label{betsrecords}

\subsection{IGUS Choice}

In this section we turn to the question of how extended probability ensemble quantum mechanics (\EP) is used. Information gathering and utilizing systems (IGUSes) like ourselves exploit the regularities summarized by physical theories to construct schemata, make predictions, and direct behavior \cite{Gel94}. Even a frog aiming to catch a fly can be said to be exploiting a rudimentary classical approximation to quantum mechanics (perhaps hard wired). The frog is in effect betting on these regularities. 

We can imagine an IGUS equipped with the \EP formulation of quantum theory of the previous section and theories of the quantum state $\Psih$ and action $\cal S$. Such an IGUS might use the theory plus data acquired in previous observations to make bets that will be settled by future data it may acquire. That is true whether or not the bet concerns the outcome of a delicate laboratory experiment, the outcome of the next election, what happened in the distant past, or whether the universe will be destroyed in a big crunch singularity in the far future.

To make a bet the IGUS must first choose the alternatives that are the basis for the bet, and then prescribe the means to settle it. The only candidates for settleable bets are sets of alternative histories that are sufficiently coarse-grained that all the $p(\alpha)$ are positive so that the set has genuine probabilities. Of course, even with that restriction there is an infinity  of possibilities.

\subsection{Records and Decoherence} 

Among the  wealth of possibilities for alternative histories to bet on  and  for ways of settling such bets, one general property stands out:  Bets are settled using records of the alternatives one is betting on.  A set of histories $\{C_\alpha\}$ is recorded if there are alternatives with projection operators $\{R_\alpha\}$ at one time that are correlated with the histories.  These alternatives need not be ones that are used or even could be used by IGUSes. As in the examples of Joos and Zeh \cite{JZ85}, there is a record of the position of a dust grain in interstellar space in all the CMB photons that have scattered from it and left the vicinity at the speed of light. 

 A very general notion of record is the following:  We say that a set of histories $\{C_\alpha\}$ is (strongly) recorded if there is an exhaustive set of orthogonal exclusive projection operators $\{R_\alpha\}$  on regions of the configuration space spanned by the $q^i$ at a time after the last alternative in $C_\alpha$   such that 
\begin{equation}
\label{records}
R_\alpha |\Psi\rangle  \approx C_\alpha |\Psi\rangle
\end{equation}
to a suitable approximation appropriate for realistic situations\footnote{There are other general notions of record. In the terminology used in \cite{Har08} we are defining strong records. There a further discussion of the appropriate degree of approximation can be found.}.    The projection $R_\alpha$ defines the record of the history $C_\alpha$. Some approximation must be allowed in \eqref{records}  because we can't expect realistic records to be {\it exactly} correlated with the histories they record. The degree of approximation is treated in \cite{Har08}. We now discuss how some familiar features of quantum mechanics are recovered   for  sets of histories $\{C_\alpha\}$ that are recorded in the sense of \eqref{records}. 

From the fundamental distribution \eqref{funddistH} and the definition of recorded sets of histories \eqref{records} it follows that 
\begin{equation}
\label{receqpos}
p(\alpha) \approx \langle \Psi |R_\alpha |\Psi\rangle = \langle \Psi |R_\alpha R_\alpha |\Psi\rangle \approx || C_\alpha |\Psi\rangle ||^2 .
\end{equation}
The extended probabilities are therefore positive for a set of histories for which there are records.  They are genuine probabilities with which to make  fair bets, and the records supply a way to settle each bet. Further, the values of the probabilities  are given by the usual quantum-mechanical square of an amplitude (the branch state vector).

A recorded set of histories is decoherent. That is because the projections $R_\alpha$ are orthogonal, so that 
\begin{equation}
\label{decoherence}
0=\langle\Psi|R_\alpha^{\dagger} R_\beta |\Psi\rangle \approx \langle \Psi |C^{\dagger}_\alpha C_\beta |\Psi\rangle \approx 0,   \quad \alpha\ne\beta ,
\end{equation}  
which is the condition for (medium) decoherence\footnote{ In a general operator context the decoherence condition implies that records of histories exist, but not necessarily in a preferred set of variables as we are assuming here. For recovering DH we need only that records imply decoherence, not the other way around.}. Any set of histories has extended probabilities, but the sets of  histories that are the basis for bets settleable by records are decoherent and have standard probabilities. 

\subsection{EPE as a version of DH}

 In equations \eqref{decoherence} and \eqref{receqpos} we have the central relations of decoherent histories quantum mechanics. Eq.\eqref{decoherence} is the condition for decoherence of the set of coarse-grained histories represented by the $\{C_\alpha\}$, and \eqref{receqpos} gives their standard probabilities. 
Thus we have recovered a sum-over-histories version of decoherent histories.  The probabilities of EPE that satisfy the sum rules of probability theory exactly are recovered only to the extent that the histories are recorded as in \eqref{records}. The DH probabilities will satisfy these rules approximately, but in realistic situations there is no physically significant difference between the two sorts of probability.  Extended probability ensemble quantum mechanics is a version of decoherent histories quantum mechanics. 

Expositions of EPE and traditional DH (TDH) begin at the same starting point and end with essentially the same predictions for probabilities of histories in realms. But the routes to these ends differ --- in particular with regard to the role played by records and decoherence. It may be helpful to the reader to compare these different routes. 

Both approaches to DH aim at  the probabilities for sets of alternative histories represented by $\{C_\alpha\}$. Both start from the observation that there are interfering sets of histories that cannot consistently be assigned probabilities. 

TDH assigns probabilities only to sets  of histories for which quantum interference between the members of the set is negligible as a consequence of the state $|\Psi\rangle$ and Hamiltonian $H$. TDH does not treat sets of histories for which interference is non-negligible. 

Decoherence means that the branch state  vectors $\{|\Psi_\alpha\rangle\}$ are nearly mutually orthogonal. This means that records of the histories exist if only trivially as projections $\{R_\alpha\}$ on the branch state vectors. The approximate probabilities of the histories are given by $p(\alpha)=||C_\alpha |\Psi\rangle||^2$. 

EPE starts from the idea that probabilities are instructions for settleable bets. But interfering sets of alternative histories are not the basis of a settleable bet. The notion of probability is expanded to give extended probabilities to non-settleable alternatives. 

Records of histories are defined as the means of settling bets on which of a set of alternative histories occurs. Decoherence is deduced as a consequence of a strong notion of records. In contrast to records following from decoherence in TDH, in EPE decoherence follows from records. The probabilities of recorded histories in EPE are $p(\alpha)=\Re\langle\Psi|C_\alpha|\Psi\rangle$, which differ negligibly from the TDH  $p(\alpha)=||C_\alpha |\Psi\rangle||^2$ in realistic situations.

\subsection{States and the Quantum-Mechanical Arrow of Time}

 The extended probability formulation of quantum theory elaborated in the previous section is time-neutral --- incorporating no fundamental arrow of time. 
 That can be seen from the fundamental distribution \eqref{funddistH} .  Field theory is invariant under $CPT$ transformations that reverse the time order of the alternatives in a $C_\alpha$.  (That is immediate in the case of chains of alternatives at moments of time, as in  \eqref{chain}.) The result is to complex conjugate the matrix elements in \eqref{funddistH} and \eqref{pfrmC}, but the resulting probabilities $p(\alpha)$ are unchanged. At this level of generality there is no quantum-mechanical arrow of time as there is in usual quantum theory and no fundamental distinction between past and future. 
  
That means that there is not a notion of a state at a moment of time that summarizes the past for prediction of the future. That is a time-asymmetric notion\footnote{See, e.g. the discussion in \cite{GH90a}. }.

However, the formula \eqref{receqpos} for the probabilities of a recorded set of histories is not time-neutral. Written out for histories that are sequences of alternatives at definite moments of time [cf \eqref{chain}] it reads
\begin{equation}
\label{chainprobs}
p(\alpha) \approx || P^n_{\alpha_n}(t_n)\cdots P^1_{\alpha_1}(t_1) |\Psi\rangle||^2 \  .
\end{equation}
The projections in the chain are time ordered\footnote{ In this paper expressions like \eqref{chainprobs} are to be understood as abbreviations for particular sum-over-histories expressions (in the preferred  variables $q^i$)  following from \eqref{fund_dist} and \eqref{cgprobs}. Thus, there is no freedom to change the times $\{t_k\}$ and thereby the descriptions of the projections (to a different set of variables using the Heisenberg equations of motion) as there is in a general operator formulation (see e.g. \cite{GH94}).}.
As already mentioned, that  is not an essential asymmetry because the order could be reversed by a $CPT$ transformation. But, in any case, there is the state at one end of the chain and nothing at the other end. This asymmetry is what is called the quantum-mechanical arrow of time\footnote{In a more general time-neutral formulation of decoherent histories quantum theory \cite{GH93b} this asymmetry can be seen as arising from an initial condition that is a pure state or a somewhat mixed state and a final condition of total indifference (a density matrix proportional to the unit matrix.) Indeed from that perspective the present formulation can be seen as one way of justifying a final condition of total indifference.}.

How did we arrive at a formulation with a quantum-mechanical arrow of time from one that did not have one? The answer is that the notion of record introduced it. To see that, first note that a consequence of \eqref{records} is
\begin{equation}
\label{corr_op}
R_\beta C_\alpha |\Psi\rangle \approx \delta_{\alpha\beta} C_\alpha|\Psi\rangle \ .
\end{equation}
The combinations $R_\beta C_\alpha$ can be considered as a set of histories in which the time of the
 records $t_{\rm rec}$  is  after the last time $t_n$ in the chain in \eqref{chainprobs}  $(t_{\rm rec}\ge t_n)$.   The result is that the probabilities are
\begin{equation}
\label{corr_prob}
p(\beta,\alpha) \approx \delta_{\beta\alpha} \ p(\alpha)
\end{equation}
which concisely expresses the correlation between records and history. With the usual conventions for past and future the $R$'s then can be  said to record the past. 

The above discussion shows that when the records defined by \eqref{records} are considered as part of history they come at the opposite end of the chain from the state $|\Psi\rangle$. That is consistent with our intuitive understanding of the processes by which records are formed in a universe like ours that has an initial low entropy state leading to a second law of thermodynamics. (See \cite{GH07} for further discussion.)

The quantum-mechanical arrow of time of DH allows us to introduce a notion of state at a moment of time as we discussed in \cite{GH90a}.  Suppose we have a decoherent set of histories consisting of sets of  $(\alpha_n, \cdots, \alpha_1)$ at sequence of times  $t_1<t_2 \cdots <t_n$ with $t_1>t_0$. Suppose that at time $t_k$  we know from present records that  particular alternatives $(\alpha_k, \cdots \alpha_1)$ happened in the past. Predictions for future alternatives are given by the conditional probabilities
\begin{equation}
\label{cond}
p(\alpha_n,\cdots,\alpha_{k+1}|\alpha_k,\cdots,\alpha_1)\equiv \frac{p(\alpha_n,\cdots,\alpha_1)}{p(\alpha_k,\cdots,\alpha_1)},
\end{equation}
where we have dropped the time labels to keep the notation concise. This can be rewritten 
\begin{align}
\label{condstate}
p(\alpha_n,&\cdots,\alpha_{k+1}|\alpha_k,\cdots,\alpha_1)\nonumber\\&\equiv ||P^n_{\alpha_n}(t_n)\cdots P^{k+1}_{\alpha_{k+1}}(t_{k+1}) |\Psi_{\alpha_k \cdots \alpha_1}\rangle ||^2 \ ,
\end{align}
where 
\begin{equation}
\label{statemomtime}
|\Psi_{\alpha_k \cdots \alpha_1}\rangle \equiv \frac{P^k_{\alpha_k}(t_k)\cdots P^1_{\alpha_1}(t_{1}|\Psi\rangle}{||P^k_{\alpha_k}(t_k)\cdots P^1_{\alpha_1}(t_{1})|\Psi\rangle||}.
\end{equation}

The state $|\Psi_{\alpha_k \cdots \alpha_1}\rangle$ defines a state at the time $t_k$ that has evolved from the begining by unitary evolution (constant in the Heisenberg picture) and projection (reduction). It summarizes past information for future prediction. We recover this time asymmetric notion of state at a moment of time, not generally, not exactly, but as a special feature of those sets of histories that are approximately recorded and are therefore  approximately decoherent\footnote{Another property of ordinary quantum mechanics that emerges only for recorded history is the usual quantum-mechanical notion of non-entangled, non-interacting subsystems represented by product states. The extended probabilities generally don't factor into products, as noted by Di\'osi \cite{Dio04}, but the probabilities for recorded histories  do, if only approximately \cite{Har08}.}.

\section{Implications of the Extended Probability Ensemble Formulation}
\label{implications}

In this section we develop the implications of the extended probability ensemble formulation of quantum theory (\EP) in which the `real' history of the universe is embedded in an ensemble of comparable imagined histories with extended probabilities. 

Throughout we will assume that histories are specified in the variables of sum-over-histories quantum mechanics (e.g. quantum fields) whether they are real or imagined, fine-grained or coarse-grained. Notions like realms (decoherent sets of coarse-grained histories) are also to be understood in this sum-over-histories context. 

A coarse graining is a partition of the histories of the ensemble into an exhaustive set of exclusive classes, which are the coarse-grained histories. The one real fine-grained history must lie in one of these classes. That class is the real coarse-grained history in the set. With sufficient and appropriate coarse graining we have realms with standard probabilities that can be the basis of settleable bets on what the real history in the set is like. One coarse-grained history in each realm is real. 

By measurement and other observation we acquire data $D$ on what the real coarse-grained history is in any realm in which $D$ is valid in some histories but not in others. With these data we also acquire coarse-grained information on what the real fine-grained history is. As we make further observations we learn more and more about the real fine-grained history. 

However, this process of progressive discovery of reality can never be carried to the completely fine-grained level. That is because the set of fine-grained histories is not decoherent, not recorded, and therefore not the basis for settleable bets\footnote{The fine-grained histories of classical statistical mechanics are only in principle the basis of a settleable bet.}.


The ensemble formulation permits the straightforward and unqualified use of ordinary language in quantum-mechanical discussion. Statements about what happened, or is happening, or will happen refer to the real history in situations where it is possible to bet on records that check these. That is, we can straightforwardly use these words in discussing realms\footnote{This is in contrast with the usual formulation of DH where it is necessary to qualify the use of `happen' by choosing to which mutually incompatible family of realms it refers \cite{Har07a}.}.

Because of the restriction to sum-over-histories variables, all coarse-grained sets are coarse grainings of the unique common fine-grained set of histories. However, even with this restriction there will generally be mutually incompatible realms (e.g as discussed in \cite{GH07}). That is, there are decoherent coarse-grained sets  for which there is not a finer-grained decoherent set of which they are both coarse-grainings \footnote{Imagine for example a closed system of two kinds of particles, red and blue. It seems likely that a set of suitably coarse-grained histories following only  the positions of the red particles could be decohered by the interactions with the blue particles. There could also be a set of coarse-grained histories following only the positions of the  blue particles are made to decohere  by interactions with the red. These realms could be mutually incompatible, with no finer-grained  decoherent set following the positions of both.}. 

{\EP is a four-dimensional presentation of quantum theory.  The histories in the ensemble are spacetime histories.  A spacelike surface of constant time particular Lorentz frame of the kind we have assumed throughout defines a notion of past, present, and future.  For any realm, and any such spacelike surface, the unique real fine-grained history selects one coarse-grained history that gives a description of what is happening in the present, what did happen in the past, and what will happen in the future. }

As is well known, spacetime formulations of theories are not inconsistent with notions of causality and IGUS-specific distinctions of past, present, and future (e.g.\cite{Har05}). The physical structure of our universe makes it easier to acquire information about times toward the big bang (what we call the past) than about times away from the big bang (what we call the future). The fact that the sources of electromagnetic radiation that we receive now lie in the past is just one of the several reasons for this \cite{Har05}. That is why there are more events in the past that can be retrodicted with high probability conditioned on present data than there are events in the future that can be predicted. 
Such time asymmetries are due to the particular quantum state of our universe\footnote{For discussions in which the authors participated, see e.g. \cite{GH93b,Har05,GH07,HHer11}.}.

\begin{table*}[t]
\caption{\bf The Ensemble in Classical Statistical Mechanics and Quantum Mechanics}
\begin{tabular}{|c||c|c|}
\hline
& Classical Statistical & Quantum Mechanics  \\   
& Mechanics & \\
\hline real   & a particular  path in phase space,  & a particular  path in configuration space,  \\
fine-grained history & $z(t)$, obeying  an equation of motion & $q(t)$ between $t_0$ and $t_f$  \\
\hline    
ensemble &alternative & alternative configuration space   \\
&phase space paths & paths between $t_0$ and $t_f$   \\
\hline     
betting  & using  probabilities & using extended probabilities --- \\
instructions& &the instruction is ``don't bet'' \\
&&if non-standard probabilities involved\\
\hline 
state & distribution on phase space & wave function \\
& $\rho(z_0,t_0) $& $\Psih(q_0,t_0)$  \\
\hline 
 fundamental &  $w[z(t)]\equiv\int dz_0 \delta[z(t)-z_t(z_0)]\rho(z_0), $ & $w[q(t)] \equiv$  \\
distribution &where $z_t(z_0)$ is the classically evolved $z_0$& $\ {\rm Re}\left[ \Psih^*(q_f,t_f) \exp\{i \Sc[q(t)]/\hbar\} \Psih(q_0, t_0) \right] \ $   \\
\hline
coarse graining &   partitions of  the ensemble into  & partitions of the ensemble into\\
& classes $c_\alpha$ (coarse-grained histories)& classes $c_\alpha$ (coarse-grained histories) \\
& one of which contains & one of which contains\\
&the real fine-grained history&the real fine-grained history\\
\hline
probabilities or & sum over fine-grained &  sum over fine-grained\\
extended probabilities& probabilities &extended probabilities \\
for coarse-grained histories & $p(\alpha) = \int_{c_\alpha} \delta z w[z(t)] $&$p(\alpha) = \int_{c_\alpha} \delta q  \ w[q(t)] $ \\
\hline   
sets that are the basis &  in principle & recorded coarse-grained sets \\
of settleable bets & all coarse-grained sets& \\
\hline           
\end{tabular}
\end{table*}

\section{Typicality Restrictions on Coarse-Grainings}
\label{restrictions}

Families of quasiclassical realms are striking features of our quantum universe. These are families of  decoherent sets of coarse-grained alternative histories defined by quasiclassical variables --- averages over suitable volumes of densities of approximately conserved quantities such as energy, momentum and particle number. The histories 
consist largely of related but branch-dependent projections onto ranges  of quasiclassical variables at a succession of times.  Each history with a non-negligible probability constitutes a narrative, with
individual histories  exhibiting patterns of correlation implied by closed sets of  effective
equations of motion interrupted by frequent small fluctuations and occasional major branchings  (as in measurement situations). 

Human IGUSes make use of coarse-grainings of a quasiclassical realm to describe the every day world of experience -- tables and chairs, stars and galaxies, measurements and their results. To be useful, a theory like that summarized in \eqref{fund_dist} must reliably describe the coarse-grained regularities of the real history in a quasiclassical realm. 

But what about other recorded coarse grainings that are allowed by the very general definition of record in \eqref{records}?
It seems likely that there are recorded sets of coarse grained histories  for which the ensemble defined by \eqref{fund_dist} will not give a reliable description\footnote{This was pointed out to us by Fay Dowker, who has worked out specific examples in simple models \cite{Dowker}.}. The allowed coarse-grained sets must therefore be more restricted than those allowed by \eqref{records}. We now describe how this works both in classical statistical mechanics and quantum mechanics. 

\subsection{Classical Statistical Mechanics}

The ensembles of the classical non-equilibrium statistical mechanics of a box of gas are usually employed to make predictions of regularities defined by quasiclassical  coarse grainings. (In the classical context we refer to these as hydrodynamic coarse grainings.)  Hydrodynamic variables are readily accessible to observation. Indeed, the probabilities of these  ensembles can be defined by maximizing the measure of ignorance (entropy) of fine-grained histories while holding fixed the ensemble averages of relevant hydrodynamic variables. Statistical mechanics makes successful predictions for such hydrodynamic coarse grainings. 

But there are coarse grainings for which the predictions of these ensembles could fail. Coarse grainings of classical histories can be defined by partitioning the space of the values of their initial phase space coordinates $z_0$. Consider the class of coarse grainings in which this initial phase space is partitioned into a small region around some particular point $z_1$ and the rest of the space. The classical ensembles will typically assign a very low probability to the small region and a probability near unity to the rest. 

Suppose the point $z_1$ happens to coincide with the initial condition $z_r$ of the real history of the gas. 
Then the ensemble will have predicted a small probability for what really will happen and a probability near unity for something else happening. 

Of course, the technology does not exist to explore such coarse grainings experimentally for a box with a large number of particles. Even it it did, it is unlikely that we would explore the one in which the real history has a small probability. Those limitations, however, should not obscure the fact that in statistical mechanics there is some  coarse graining for which the real history has negligible probability. The probabilities assigned by the ensemble do not, in this sense, constitute a good description of the real history. 

If such coarse grainings are not accessible by realistic experiment then the theory can perhaps be augmented by restrictions on the coarse-grainings for which it can be employed --- to quasiclassical realms for instance. But if observations can be carried out to decide which history in such a coarse graining actually occurs then we have to allow that the observed (real) history may be one that was assigned a very low probability. 

A characteristic feature of such coarse grainings is that the real coarse-grained history is not a typical history of the ensemble. More precisely, suppose $\{p(\alpha)\}$ are the probabilities assigned by the ensemble to the alternative coarse-grained histories; let us call  the real coarse-grained history $r$. Then the following typicality condition \cite{GL04} is not satisfied:
\begin{equation}
\label{typicality}
-\log p(r) \ll -\sum_\alpha p(\alpha) \log p(\alpha) = S \ ,
\end{equation} 
where $S$ is the entropy of the ensemble's probability distribution. 

 Would we discard statistical mechanics if such an experiment were carried out with this result? It would be more sensible to restrict the coarse grainings that  test the theory to ones in which the typicality condition \eqref{typicality} is satisfied. Such coarse grainings cannot be identified in advance of the determination of the real coarse-grained history. 
But typicality can be calculated when the real results are in and their significance as a test of the theory is assessed accordingly. 

\subsection{Quantum Mechanics}
The situation in the EPE formulation of quantum mechanics is similar. For any fine-grained history it is possible to find a coarse graining in which its probability is very small or even zero. For example, consider a partition of the fine-grained histories $q(t)$ by whether the values of $q$ at a time $t_1$ are in a small region around $q_1$ or elsewhere. Unless the state and dynamics are very special, the fundamental distribution \eqref{fund_dist} will assign a small probability to the small region. If $q_1$ happened to coincide with the $q$ of the real history at that time the real history would be assigned a small or zero probability\footnote{Indeed, Dowker has exhibited \cite{Dowker} simple examples involving spin in which, for every fine-grained history, there is some coarse graining in which its probability is zero.}. 


The notion of record in \eqref{records} that implies medium decoherence is general and mathematically simple. The quasiclassical coarse grainings used by human IGUSes are a much more restricted class. Restrictions on either the coarse-grainings or their records are can still lead to  theories that are consistent with known observations and would rule out examples such as those discussed above. 

 Whatever the restriction on records and coarse grainings, one should check  that the real coarse-grained history that emerges has a probability that is typical of the ensemble of coarse-grained histories before throwing out the theory in \eqref{fund_dist}, which consists of  $\Psih$, ${\cal S}$, and quantum mechanics itself.

\section{Discussion}
\label{conclusion}

By way of conclusion, let us begin by recapitulating the analogy between classical statistical mechanics (CSM) and extended probability ensemble quantum theory with one real history (\EP). 

There is a common central idea in CSM and \EP:  They are both concerned with systems with one real fine-grained history about which we have little information from observation either in practice (CSM) or in principle (\EP). The coarse-grained regularities that are accessible to observation and test cannot therefore be predicted from a fine-grained starting point. Rather, both theories use the ensemble method to describe coarse-grained regularities. The real history is conceptually embedded  in an ensemble of comparable imagined histories that are assigned measures of ignorance consistent with coarse-grained knowledge\footnote{ Gibbs considered a real system (box of gas) embedded in an ensemble of imagined systems. In what amounts to the same thing, we are considering an ensemble of real and imagined histories of one system.}.  Within this ensemble framework there are similarities and differences between CSM and \EP.  The table summarizes these, but we now describe them in a few additional words.  

{\it Notion of probability:}  In CSM the fine-grained history is assumed to be determinable in principle and therefore the basis of a settleable bet on what it is like. Standard probabilities are therefore assigned to members of the ensemble. In \EP the fine-grained history is in principle not completely accessible  and therefore not the basis of a settleable bet. The ensemble therefore has to be constructed with extended probabilities. 

{\it Assignment:} In CSM, for the case of local equilibrium, the probabilities can be constructed by a principle of maximizing entropy, the measure of ignorance, while holding certain quantities fixed. (Energy, momentum, and particle number in suitably small volumes are examples.)
In \EP the extended probabilities are given as the fundamental distribution \eqref{fund_dist} based on  the state and Hamiltonian. Whether there are deeper principles that underlie this formula is an open question.

Despite these differences the two theories are similar in their general characteristics. In particular, \EP can be regarded as a classical stochastic theory based on extended probabilities. That perspective has a number of advantages for interpreting quantum mechanics that we have discussed earlier: \EP provides a unified perspective on coarse-graining. \EP allows the use of ordinary language especially with existential words such as `happen'.  \EP provides a simple interpretation of extended probabilities as a unified measure both of ignorance and knowability in principle. { The sum rules of probability theory are satisfied exactly for the extended probabilities of \EP.} \EP may provide a starting point for further generalization and modification of quantum mechanics. But perhaps most interestingly \EP provides an intuitively attractive notion of reality in quantum theory.

If a notion of reality is to be introduced in DH it seems only natural that there be one real history in each realm. The question of the connection between the real histories in different realms then arises\footnote{There is a significant literature discussing reality in the context of decoherent histories quantum mechanics. Some recent discussions can be found in \cite{Gri11,MQW,Hoh11} with references to further literature.}.
Two realms may be compatible in the sense that there is a finer-grained realm of which they are both coarse grainings. In that case the real history in each of the coarser-grainded realms is the one that contains the real history of the finer-grained realm. 
However, two realms may be incompatible --- without a finer-grained realm of which they are both coarse grainings. Unadorned DH provides no connection between the meanings of  reality for incompatible realms. This poses a challenge to the notion of reality in unadorned DH. 

This challenge is overcome in the \EP version of DH. There is one real fine-grained history in an ensemble of comparable histories with extended probabilities. All realms are coarse grainings of this unique fine-grained set described in sum-over-histories variables. The  real history in each realm is the one that contains the real fine-grained history. Thus \EP has a single notion  of reality that is expressed in all realms.

{\acknowledgments
We thank Fay Dowker for a helpful exchange. JBH thanks the Santa Fe Institute for supporting many visits there. The work of JBH was also supported in part by the National Science Foundation under grant PHY07-57035.
MG-M and JBH acknowledge the support of the Bryan J. and June B. Zwan Foundation. MG-M was in addition supported by the C.O.U.Q. Foundation,  by Insight Venture Partners, and by the KITP in Santa Barbara.  The generous help provided by these organizations is gratefully acknowledged by both authors.}
 
 \section*{APPENDIX: QUESTIONS THAT MIGHT BE FREQUENTLY ASKED}

\renewcommand{\theequation}{\Alph{section}.\arabic{equation}}

{\bf Bell's Inequalities:}  {\it Since EPE can be thought of as a classical stochastic theory, won't it imply that the Bell inequalities are  satisfied? } 
{ As is well known \cite{Bell}, to understand classically the experiments that demonstrate a violation of Bell's inequalities requires either non-local interactions or negative probabilities or both. EPE can be viewed as a classical stochastic theory {\it with extended probabilities that can sometimes be negative.}  There is thus no conflict with these experiments. Indeed, EPE is a formulation of quantum theory, so it predicts a violation of the Bell inequalities.}

{\bf Hidden Variables:} {\it Is this in effect a hidden variable theory?}   There are no variables involved beyond the usual quantum fields of sum-over-histories quantum theory --- the $\{q(t)\}$. However their fine-grained values are not completely accessible to experiment or observation and therefore partially hidden. 

{\bf One real history in each realm:} {\it  Why are extended probabilities needed? Couldn't one say that in DH one history in each realm is real?} {  One can say that. But, as discussed in Section \ref{conclusion}, using extended probability for  fine-grained histories provides a connection between the realities of different realms.}

{\bf Probabilities and Frequencies:} {\it I understand probabilities as frequencies, but a negative number cannot be a frequency.}  We are taking a more general view of probabilities,   commonly called a Bayesian one,  of probabilities as instructions for betting.  Extended probabilities outside the range \range\ are, in a way,  instructions for betting --- don't bet, it won't be settleable.  The connection of standard probabilities with frequencies of occurrences in an infinite ensemble of identical systems can be derived from this view of probability {for realms.} See e.g. \cite{PROBS}.

\end{document}